\documentclass[aps,prl,twocolumn,superscriptaddress,bibnotes,floatfix,preprintnumbers]{revtex4-1}
\pdfoutput=1
\usepackage{graphicx}
\usepackage{bm}
\usepackage{amssymb,amsmath,latexsym}
\usepackage{color}
\usepackage{lineno}
\usepackage{comment}
\definecolor{darkblue}{RGB}{0,0,196}
\definecolor{darkred}{RGB}{196,0,0}
\definecolor{darkgreen}{RGB}{0,136,0}
\usepackage[colorlinks=true,linktocpage=true,linkcolor=darkblue,citecolor=darkblue,urlcolor=darkblue]{hyperref}
\usepackage{array}
\usepackage{soul}

\newcommand{\mes}{\Delta}

\begin{document}
%\linenumbers

\title{Regeneration of bottomonia in an open quantum systems approach}

\preprint{TUM-EFT 178/23; FERMILAB-PUB-23-060-V}

\author{Nora Brambilla}
%\email{nora.brambilla@ph.tum.de}
\affiliation{Physik-Department, Technische Universit\"{a}t M\"{u}nchen, James-Franck-Str. 1, 85748 Garching, Germany}
\affiliation{Institute for Advanced Study, Technische Universit\"{a}t M\"{u}nchen, Lichtenbergstrasse 2 a, 85748 Garching, Germany}
\affiliation{Munich Data Science Institute, Technische Universit\"{a}t M\"{u}nchen, Walther-von-Dyck-Strasse 10, 85748 Garching, Germany}

\author{Miguel \'{A}ngel Escobedo}
%\email{miguelangel.escobedo@usc.es}
\affiliation{Instituto Galego de F\'{i}sica de Altas Enerx\'{i}as (IGFAE), Universidade de Santiago de Compostela. E-15782, Galicia, Spain}
\affiliation{Departament de Física Quàntica i Astrofísica and Institut de Ciències del Cosmos, Universitat de Barcelona, Martí i Franquès 1, 08028 Barcelona, Spain}

\author{Ajaharul Islam}
%\email{aislam2@kent.edu}
\affiliation{Department of Physics, Kent State University, Kent, Ohio 44242, United States}

\author{Michael Strickland}
%\email{mstrick6@kent.edu}
\affiliation{Department of Physics, Kent State University, Kent, Ohio 44242, United States}

\author{Anurag Tiwari}
%\email{anurag.tiwari128@gmail.com}
\affiliation{Department of Physics, Kent State University, Kent, Ohio 44242, United States}

\author{Antonio Vairo}
%\email{antonio.vairo@tum.de}
\affiliation{Physik-Department, Technische Universit\"{a}t M\"{u}nchen, James-Franck-Str. 1, 85748 Garching, Germany}

\author{Peter Vander Griend}
%\email{vandergriend@tum.de}
\affiliation{Department of Physics and Astronomy, University of Kentucky, Lexington, Kentucky 40506, United States}
\affiliation{Theoretical Physics Department, Fermilab, P.O. Box 500, Batavia, Illinois 60510, United States}

\begin{abstract}
We demonstrate the importance of quantum jumps in the nonequilibrium evolution of bottomonium states in the quark-gluon plasma. Based on nonrelativistic effective field theory and the open quantum system framework, we evolve the density matrix of color singlet and octet pairs. We show that quantum regeneration of singlet states from octet configurations is necessary to understand experimental results for the suppression of both bottomonium ground and excited states. The values of the heavy-quarkonium transport coefficients used are consistent with recent lattice QCD determinations. 
\end{abstract}

\date{\today}


\maketitle

The suppression of bottomonium production in heavy-ion collisions relative to proton-proton collisions provides strong evidence of the formation of a hot, deconfined quark-gluon plasma (QGP)~\cite{Acharya:2020kls,ATLAS5TeV,Sirunyan:2018nsz,CMS:2017ycw,CMS:2020efs,ALICE:2019pox,STAR:2013kwk,PHENIX:2014tbe,STAR:2016pof,CMS-PAS-HIN-21-007,STAR:2022rpk}. 
Studies of quarkonium suppression were triggered in 1986 by 
the idea of Matsui and Satz relating suppression to the Debye screening of interactions in a color-ionized QGP~\cite{Matsui:1986dk}.
In recent years, detailed quantum chromodynamics (QCD) calculations have demonstrated that in-medium heavy-quarkonium  suppression is due to two effects:
screening and the generation of a large imaginary part of the in-medium heavy-quarkonium potential, with the latter effect dominating over the former in the temperature regime of interest~\cite{Laine:2006ns,Brambilla:2008cx,Beraudo:2007ky,Escobedo:2008sy,Dumitru:2009fy,Brambilla:2010vq,Brambilla:2011sg,Brambilla:2013dpa}.  Although this new understanding represents a fundamental paradigm shift, it has further reinforced the importance of heavy-quarkonium suppression as an ideal probe of the QGP.

Prior studies \cite{Brambilla:2016wgg,Brambilla:2017zei} have shown that the in-medium quantum evolution of heavy quarkonium depends on the heavy-quarkonium momentum diffusion coefficient $\hat{\kappa}$ and its dispersive counterpart $\hat{\gamma}$, with both defined in QCD in terms of nonperturbative correlators of chromoelectric fields.
In the context of the in-medium evolution equations, $\hat{\kappa}$ and $\hat{\gamma}$ appear as in-medium corrections to the imaginary and real parts of the heavy quarkonium potential, respectively.
In this work, we present results obtained using state of the art determinations of $\hat{\kappa}$ \cite{Larsen:2019bwy,Bala:2021fkm} and $\hat{\gamma}$ \cite{Bala:2021fkm} from unquenched lattice QCD measurements. 
In contrast with earlier determinations \cite{Brambilla:2019tpt,Aarts:2011sm,Kim:2018yhk}, these recent measurements tend to favor somewhat larger values of $\hat{\kappa}$ and a thermal part of $\hat{\gamma} \approx 0$~\footnote{ 
We find, intriguingly, that the phenomenological heavy quarkonium suppression results reported in this work favor a value of $\hat{\kappa}$ that is consistent with its most recent lattice determinations and a value of $\hat{\gamma}$ around zero.  We also remark that recent lattice studies \cite{Altenkort:2023oms} of the heavy-quark momentum diffusion coefficient (a quantity related to $\hat{\kappa}$ but in the fundamental rather than adjoint representation) have found larger values than previous calculations \cite{Meyer:2010tt,Banerjee:2011ra,Francis:2015daa,Brambilla:2020siz,Altenkort:2020fgs}.
%We also remark that the lattice studies~\cite{Larsen:2019bwy} and~\cite{Bala:2021fkm} give information on $\hat{\kappa}$ and  $\hat{\gamma}$ in the adjoint representation, which are the relevant quantities for our study.
}.
A combination of a large in-medium width, related to a large  $\hat{\kappa}$, and a small thermal mass shift, related to a small thermal part of $\hat{\gamma}$, provide further support for changing the original screening picture of heavy quarkonium suppression, as proposed by Matsui and Satz, into a suppression mechanism dominated by in-medium dissociation due to the thermal width.

Effective field theories (EFTs), specifically potential  nonrelativistic quantum chromodynamics (pNRQCD) \cite{Pineda:1997bj,Brambilla:1999xf,Brambilla:2004jw}, allow us to systematically exploit the hierarchies of energy scales of the quarkonium system and the thermal medium.  Based on these hierarchies, one can factorize 
the contributions from the different scales and thus 
significantly simplify the problem, while still performing the calculations in QCD  
in a fully quantum setting.
Furthermore, the formalism of open quantum systems (OQS) \cite{Breuer:2002pc} allows us to rigorously treat a quantum system that evolves coupled to and out of equilibrium with a thermal bath.
Making use of these two theoretical tools, we are able to describe the evolution of the system, fully taking into account its quantum non-Abelian nature and the out-of-equilibrium evolution~\footnote{
In~\cite{Brambilla:2022ynh}, we obtained a reasonable description 
of the bottomonium nuclear modification factor by neglecting 
some quantum effects (quantum jumps).
However, increased statistics and new observables from heavy-ion collision experiments on bottomonium suppression call for a full quantum description, as we shall argue in the rest of the Letter.
}.

In the remainder of this Letter, we present our method, discussing pNRQCD and the heavy-quarkonium evolution equations, and results including the nuclear modification factor $R_{AA}$ of the $\Upsilon(1S)$, $\Upsilon(2S)$, and $\Upsilon(3S)$ (and double ratios thereof) with respect to the number of participating nucleons $N_{\rm part}$ and transverse momentum $p_{T}$.
We compare our results to measurements of the ALICE \cite{Acharya:2020kls}, ATLAS \cite{ATLAS5TeV}, and CMS \cite{Sirunyan:2018nsz,CMS-PAS-HIN-21-007} Collaborations and observe good agreement with the experimental data for all observables.
Our main findings are twofold: namely that the experimental data cannot be reproduced in the absence of quantum regeneration and, in accordance with recent lattice measurements, a larger in-medium width and smaller mass shift, i.e., larger $\hat{\kappa}$ and smaller $|\hat{\gamma}|$, than previously measured are favored by current bottomonium suppression results.

{\em Methodology} -- Heavy quarkonium states are characterized by separated energy scales making them ideal for description using EFTs.
Heavy-heavy bound states are characterized by the heavy quark mass $M$ and the nonrelativistic relative velocity $v\ll 1$.
Within this context, $M$, $Mv$, and $Mv^{2}$ are referred to as the hard, soft, and ultrasoft scales, respectively.
Integrating out the hard scale from  QCD gives rise to nonrelativistic QCD (NRQCD) \cite{Caswell:1985ui,Bodwin:1994jh}.
Use of NRQCD to describe heavy-heavy bound states is problematic as both the soft scale characterizing the momentum transfer and the ultrasoft scale corresponding to the binding energy $E$, remain dynamical, and no unambiguous power counting can be assigned to the operators appearing in the NRQCD Lagrangian without additional assumptions.
This issue is remedied by integrating out the soft scale giving rise to potential NRQCD (pNRQCD)
\cite{Pineda:1997bj,Brambilla:1999xf,Brambilla:2004jw}.

pNRQCD implements an expansion in the inverse of the heavy quark mass $M^{-1}$ and the bound-state radius $r\sim (Mv)^{-1}$ at the Lagrangian level and is thus ideally suited to describe small, heavy-heavy bound states.
After integrating out the soft scale, the individual heavy quarks are no longer resolved, and the degrees of freedom in the pNRQCD Lagrangian are heavy-heavy bound states in color singlet and color octet configurations and gluons and light quarks at the ultrasoft scale.
Matching pNRQCD onto NRQCD gives rise to non-local potentials 
at lowest order in the power counting.
If we assume that the radius of the system is smaller than $1/\Lambda_\textrm{QCD}$, then the potentials are the 
attractive and repulsive Coulombic potentials in the singlet and octet sectors, respectively. This assumption may apply to the lowest bottomonium states.

In Refs.~\cite{Brambilla:2016wgg,Brambilla:2017zei} the authors used pNRQCD and the OQS formalism to derive a master equation for a heavy, Coulombic quarkonium realizing the following 
hierarchy of scales in a strongly coupled ($T\sim g T$)  thermal QCD medium (the QGP)
\begin{equation}
	M \gg 1 / a_{0} \gg \pi T ,
\end{equation}
where $a_{0}$ is the Bohr radius of the bound state.
They derived, furthermore, a Lindblad equation~\cite{Gorini:1975nb,Lindblad:1975ef} valid in the additional limit $\pi T \gg E$.
References~\cite{Brambilla:2020qwo,Brambilla:2021wkt} solved this Lindblad equation by developing an open source code, called QTraj \cite{Omar:2021kra}, which implements the quantum trajectories algorithm \cite{Daley:2014fha}.
Reference~\cite{Brambilla:2022ynh} used the QTraj code to solve a Lindblad equation accurate up to and including terms of order $E/(\pi T)$ allowing for an extension of the region of validity to a lower final temperature $T_F$ nearer the pseudocritical temperature $T_{\rm pc}$ of the QGP phase transition.
The next-to-leading order (NLO) Lindblad equation is 
\begin{widetext}
\begin{equation}
	\frac{d\rho(t)}{dt} = -i \left[ H, \rho(t) \right] + \sum_{n=0}^1
	\left( C_{i}^{n} \rho(t) C^{n\dagger}_{i} - \frac{1}{2} \left\{ C^{n \dagger}_{i} C_{i}^{n}, \rho(t) \right\} \right),
	\label{eq:Lindblad}
\end{equation}
where $\rho(t)$ and $H$ represent the quarkonium density matrix and Hamiltonian including in-medium corrections
\begin{equation}
	\rho(t) = \begin{pmatrix} \rho_{s}(t) & 0 \\ 0 & \rho_{o}(t) \end{pmatrix}, \quad
	H = \begin{pmatrix} h_{s} + \frac{r^{2}}{2} \hat{\gamma}\, T^{3} +\frac{\hat{\kappa}T^{3}}{4MT} \{r_{i}, p_{i}\} & 0 \\ 0 & h_{o} + \frac{N_{c}^{2}-2}{2(N_{c}^{2}-1)} \left( \frac{r^{2}}{2} \hat{\gamma}T^{3} +\frac{\hat{\kappa}T^{3}}{4MT} \{r_{i}, p_{i}\} \right) \end{pmatrix},
\end{equation}
and the $C_{i}^{n}$ are collapse operators encoding the in-medium width
\begin{align}
	&\begin{aligned}\label{eq:c0}
		C_{i}^{0} =& \sqrt{\frac{\hat{\kappa}T^{3}}{N_{c}^{2}-1}} \begin{pmatrix} 0 & 1 \\ 0 & 0 \end{pmatrix} \left(r_{i} + \frac{i p_{i}}{2MT} +\frac{\Delta V_{os}}{4T}r_{i} \right) + \sqrt{\hat{\kappa}T^{3}} \begin{pmatrix} 0 & 0 \\ 1 & 0 \end{pmatrix} \left(r_{i} + \frac{i p_{i}}{2MT} +\frac{\Delta V_{so}}{4T}r_{i} \right),
	\end{aligned}\\
	&C_{i}^{1} = \sqrt{\frac{\hat{\kappa}T^{3}(N_{c}^{2}-4)}{2(N_{c}^{2}-1)}} \begin{pmatrix} 0 & 0 \\ 0 & 1 \end{pmatrix} \left(r_{i} + \frac{i p_{i}}{2MT} \right).\label{eq:c1}
\end{align}
\end{widetext}
$\rho_{s,o}(t)$ and $h_{s,o}(t)$ represent the heavy quarkonium singlet, octet density matrix and vacuum Hamiltonian; $r_{i}$ and $p_{i}$ are the position and momentum operators, respectively, associated with the radial coordinate; and $\Delta V_{so}=-\Delta V_{os}=V_{s}-V_{o}$ is the difference of the singlet and octet potentials.
For further details on the above Lindblad equation and its derivation we refer the reader to Refs.~\cite{Akamatsu:2020ypb,Brambilla:2022ynh}.

The heavy quarkonium momentum diffusion coefficient $\hat{\kappa}$ and its dispersive counterpart $\hat{\gamma}$ are defined as
\begin{align}
	\hat{\kappa} &= \frac{1}{T^{3}}\frac{g^{2}}{6N_{c}} \int_{0}^{\infty}\text{d}s\, \left\langle \left\{\tilde{E}^{a}_i(s,\vec{0}), \tilde{E}^{a}_i(0,\vec{0})\right\}\right\rangle ,\label{eq:kappa}\\ 
	\hat{\gamma} &= -\frac{i}{T^{3}}\frac{g^{2}}{6N_{c}} \int_{0}^{\infty}\text{d}s\, \left\langle \left[\tilde{E}^{a}_i(s,\vec{0}), \tilde{E}^{a}_i (0,\vec{0})\right]\right\rangle,\label{eq:gamma}
\end{align}
where $\tilde{E}^{a}_i$ is a chromoelectric field with Wilson lines attached (see Eqs.~(2.9) and (2.10) of Ref.~\cite{Brambilla:2022ynh}) and $N_{c}=3$ is the number of colors.
We note that the above transport coefficients are in the \textit{adjoint} representation as
 obtained in Refs.~\cite{Brambilla:2016wgg,Brambilla:2017zei,Eller:2019spw}. They are given in terms of  nonperturbative correlators that should be evaluated in QCD with nonperturbative methods   and contain the information about heavy-quarkonium coupling to the QGP.

For the simulation details used to solve the Lindblad equation
including temporal and spatial discretizations, initial state, etc., see 
the introduction of Sec.~4 of Ref.~\cite{Brambilla:2022ynh}; results in the present work differ only in the values of the transport coefficients $\hat{\kappa}$ and $\hat{\gamma}$ and the inclusion of quantum jumps, i.e., quantum regeneration.

In order to accurately describe the in-medium, nonequilibrium evolution of heavy quarkonium, one must include dissociation and recombination.
The Lindblad equation accounts for these processes via the collapse operators given in Eqs.~\eqref{eq:c0} and \eqref{eq:c1}, which implement transitions among the different color and angular momentum states.
The Hamiltonian and the anticommutator terms in Eq.~\eqref{eq:Lindblad} preserve the quantum numbers of the state while reducing its norm; the first term in parentheses in Eq.~\eqref{eq:Lindblad} changes the quantum numbers of the system and ensures that the overall evolution is trace preserving.
The former terms implement dissociation while the latter term implements \textit{quantum jumps} which are responsible for quantum recombination.
Due to computational costs, in our most recent work \cite{Brambilla:2022ynh}, we implemented only dissociation and observed reasonable agreement with experimental measurements of the nuclear modification factor for the ground state.
In this work, we include the effect of quantum jumps and observe this effect to be of critical importance for simultaneously describing the suppression of both the ground and excited bottomonium states.

{\em Results} -- We performed numerical simulations with and without quantum jumps for $\hat\kappa \in \{2,3,4,5\}$ and $\hat\gamma \in \{-3.5,-2.6,0,1\}$.  
For the background evolution, we made use of previously generated (3+1)-D anisotropic hydrodynamics backgrounds \cite{Alqahtani:2020paa} and initialized the QTraj Lindblad equation solver at $\tau_{\rm med} = 0.6$~fm\,\footnote{For details concerning the hydrodynamic temperature evolution see Refs.~\cite{Brambilla:2020qwo,Brambilla:2021wkt,Brambilla:2022ynh} wherein the same hydrodynamic evolution \cite{Alqahtani:2020paa} was used.}\footnote{For details concerning the quantum trajectories method used see Refs.~\cite{Omar:2021kra,Brambilla:2022ynh}.}.  
The sampling of initial production points and transverse momenta was performed in the same manner as our prior works~\cite{Brambilla:2020qwo,Brambilla:2021wkt,Brambilla:2022ynh}. 
We did not include initial state energy loss or other cold nuclear matter effects. 

In all cases discussed herein, we generated approximately 100,000{\,-\,}200,0000 physical trajectories. 
When including quantum jumps, we generated approximately 30 quantum trajectories per physical trajectory \footnote{The term quantum trajectory has to be understood in the context of the Quantum Trajectory Method used to solve the Lindblad equation \cite{Dalibard:1992zz}. Physical trajectories are the trajectories that quarkonium follows in physical space according to the initial conditions.}.   
We initialized the quantum wave function as a 
highly-localized Gaussian delta function 
%following our prior works 
\cite{Brambilla:2020qwo,Brambilla:2021wkt,Brambilla:2022ynh,Alalawi:2022gul} and terminated the evolution along each physical trajectory when the temperature dropped below $T_F = 190$ MeV \footnote{This lower temperature cutoff ensures that the NLO corrections to the bottomonium decay widths remain less than 50\%, as discussed in Ref.~\cite{Brambilla:2022ynh}.}.
We computed the survival probabilities of the $\Upsilon(1S,2S,3S)$ and $\chi_b(1P,2P)$ states by taking the ratio of the final and initial overlap probabilities obtained using vacuum Coulomb eigenstates.  
We then took into account final state feed down by applying a feed down matrix constructed from data available from the Particle Data Group \cite{Workman:2022ynf}.
The feed down matrix used was the same as in our prior works \cite{Brambilla:2020qwo,Brambilla:2021wkt,Brambilla:2022ynh,Alalawi:2022gul}.

\begin{figure}[t]
    \centering
    \includegraphics[width=0.98\linewidth]{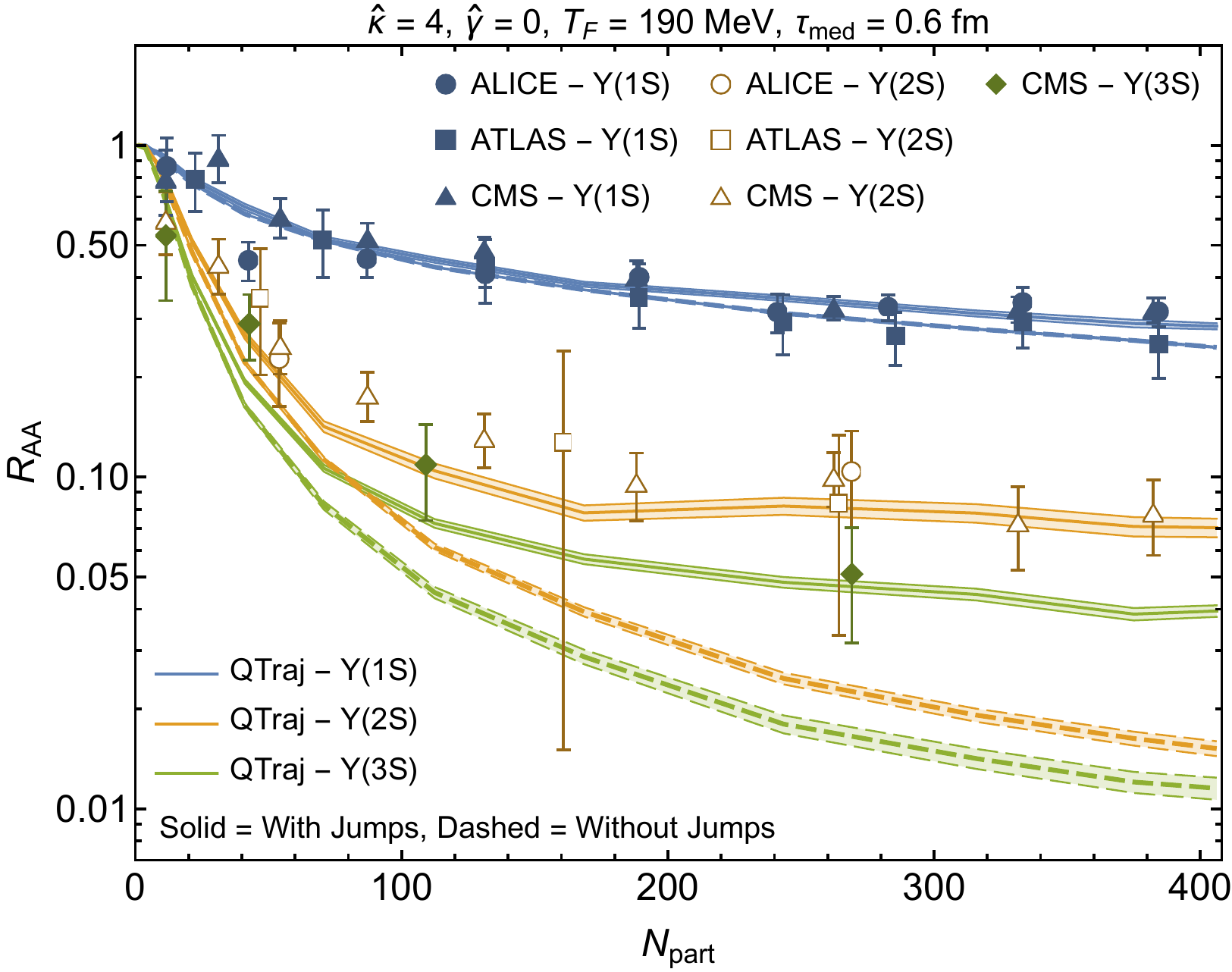}
    \vspace{-3mm}
    \caption{The nuclear suppression factor $R_{AA}$ of bottomonia states as a function of $N_{\rm part}$.  The solid and dashed lines indicate results obtained with and without quantum jumps, respectively.  The experimental measurements shown are from the ALICE~\cite{Acharya:2020kls}, ATLAS~\cite{ATLAS5TeV}, and CMS~\cite{Sirunyan:2018nsz,CMS-PAS-HIN-21-007} Collaborations.}
    \label{fig:raavsnpart-log}
\end{figure}

\begin{figure}[t]
    \centering
    \includegraphics[width=0.98\linewidth]{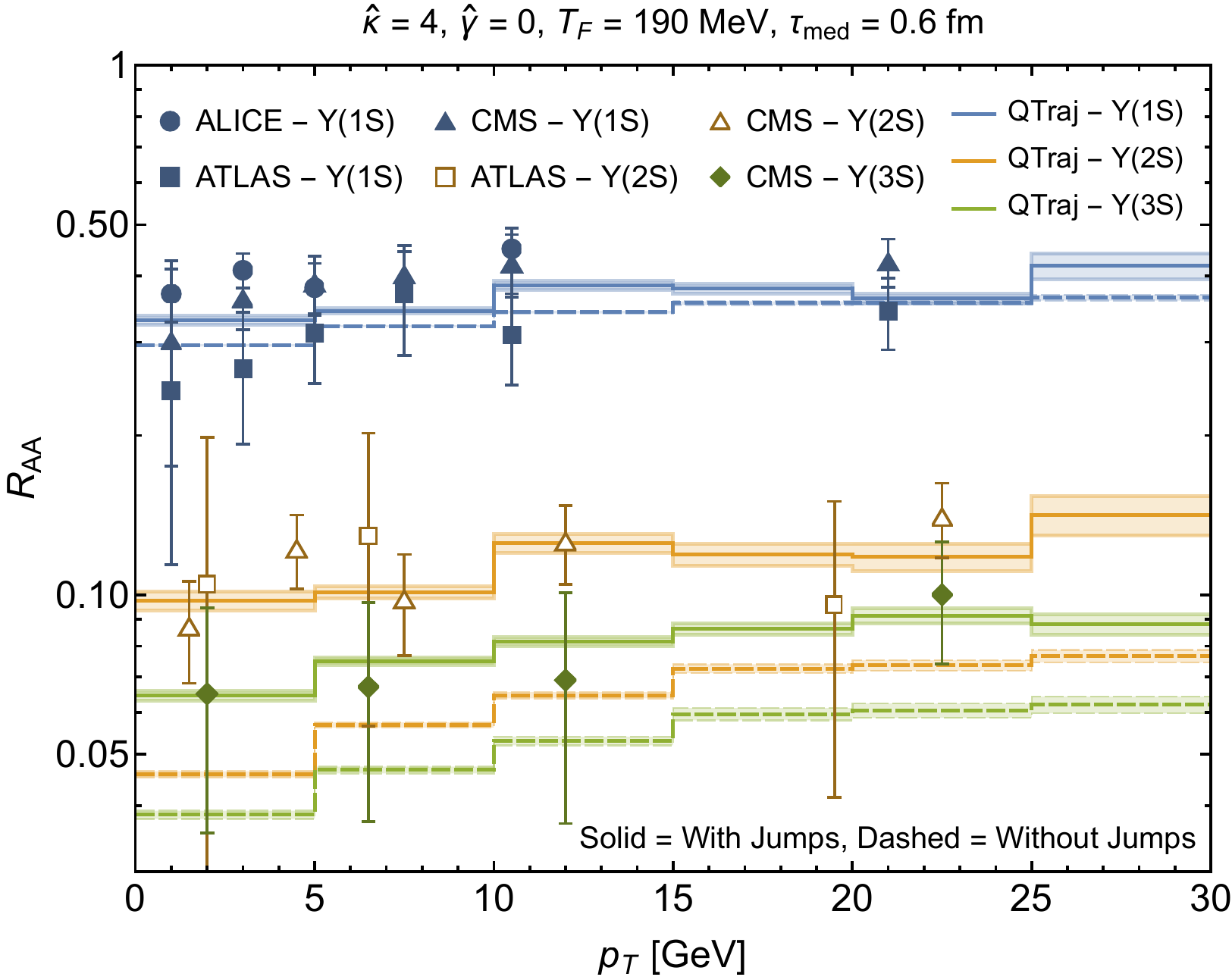}
    \vspace{-3mm}
    \caption{$R_{AA}$ of bottomonia states as a function of $p_T$.  Line styles and experimental data sources are the same as Fig.~\ref{fig:raavsnpart-log}. }
    \label{fig:raavspt-log}
\end{figure}

We analyzed the agreement of the resulting NLO QTraj predictions for $R_{AA}(N_{\rm part})$ with data available from the ALICE, ATLAS, and CMS experiments and found that, without including quantum jumps, it was impossible to simultaneously describe the suppression of the $\Upsilon(1S)$ and $\Upsilon(2S)$ states.  
Contrarily, when quantum jumps were included, we found that it was possible to quantitatively describe the suppression of both states.  
In addition, we found that this allowed for a good description of the suppression of the $\Upsilon(3S)$.  
For details concerning the measure we used to assess agreement with experimental data, we refer the reader to the supplemental material associated with this Letter.  
Therein, we present plots of the agreement measure as a function of $\hat\kappa$ and $\hat\gamma$.  
From our analysis, we concluded that the best description of the data was obtained using the values $\hat\kappa \approx 4$ and $\hat\gamma \approx 0$ from our original set of simulations.  
In all figures, we present results obtained with these values of $\hat\kappa$ and $\hat\gamma$ and assess the effect of including quantum regeneration.

In Figs.~\ref{fig:raavsnpart-log} and \ref{fig:raavspt-log}, we present a comparison of the NLO QTraj results and experimental data as functions of $N_{\rm part}$ and transverse momentum, respectively.  
The vertical axis in both plots is logarithmic in order to better resolve the suppression of the excited states.  
In these figures and all subsequent figures, the solid and dashed lines correspond to the results obtained with and without quantum jumps, respectively.  
The shaded bands indicate the statistical uncertainty associated with the average over quantum and physical trajectories.  
As can be seen from Figs.~\ref{fig:raavsnpart-log} and \ref{fig:raavspt-log}, without quantum jumps included, one under predicts $R_{AA}$ for excited states.  
This is not particular to the case of $\hat\kappa = 4$ and $\hat\gamma = 0$ and was observed in our prior simulations in which a temperature-dependent $\hat\kappa$ was used \cite{Brambilla:2020qwo,Brambilla:2021wkt,Brambilla:2022ynh,Alalawi:2022gul}.  
When quantum jumps are included, we find excellent agreement with the experimentally measured suppression of both the ground and excited states.
For the fully integrated $R_{AA}[1S,2S,3S]$, we obtain $0.3598 \pm 0.0028$, $0.1118 \pm 0.0019$, and $0.0775 \pm 0.0007$, respectively.

\begin{figure}[t]
    \centering
    \includegraphics[width=0.98\linewidth]{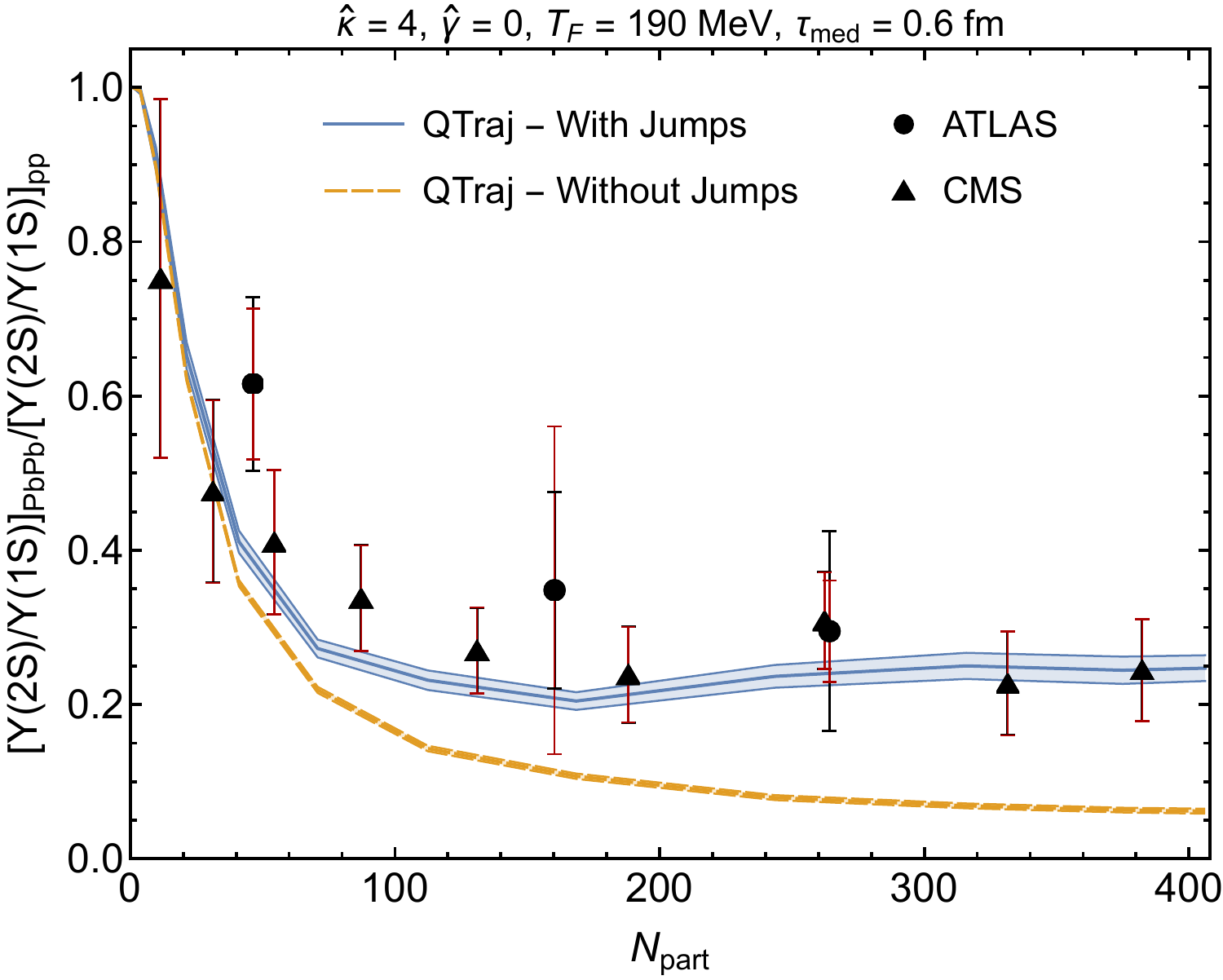}
    \vspace{-4mm}
    \caption{The 2S/1S double ratio as a function of $N_{\rm part}$.  The solid blue lines and dashed orange lines show the QTraj results with and without quantum jumps, respectively. The experimental measurements shown are from the ATLAS~\cite{ATLAS5TeV} and  CMS~\cite{Sirunyan:2018nsz,CMS-PAS-HIN-21-007} Collaborations.}
    \label{fig:ratio2s1svsnpart}
\end{figure}

\begin{figure}[t]
    \centering
    \includegraphics[width=0.98\linewidth]{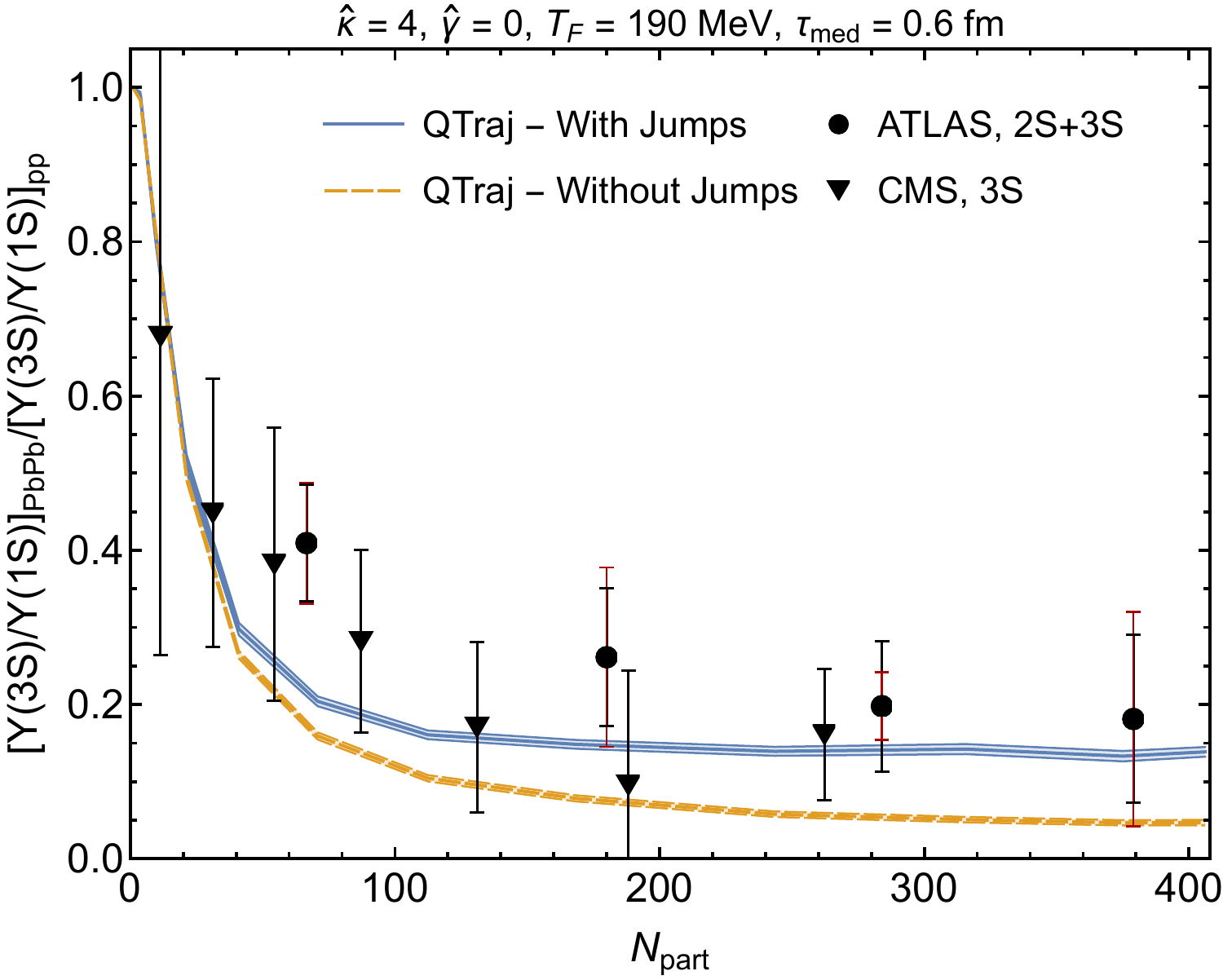}
    \vspace{-4mm}
    \caption{The 3S/1S double ratio as a function of $N_{\rm part}$.  Line styles and experimental data sources are the same as Fig.~\ref{fig:ratio2s1svsnpart}.}
    \label{fig:ratio3s1svsnpart}
\end{figure}

\begin{figure}[t]
    \centering
    \includegraphics[width=0.98\linewidth]{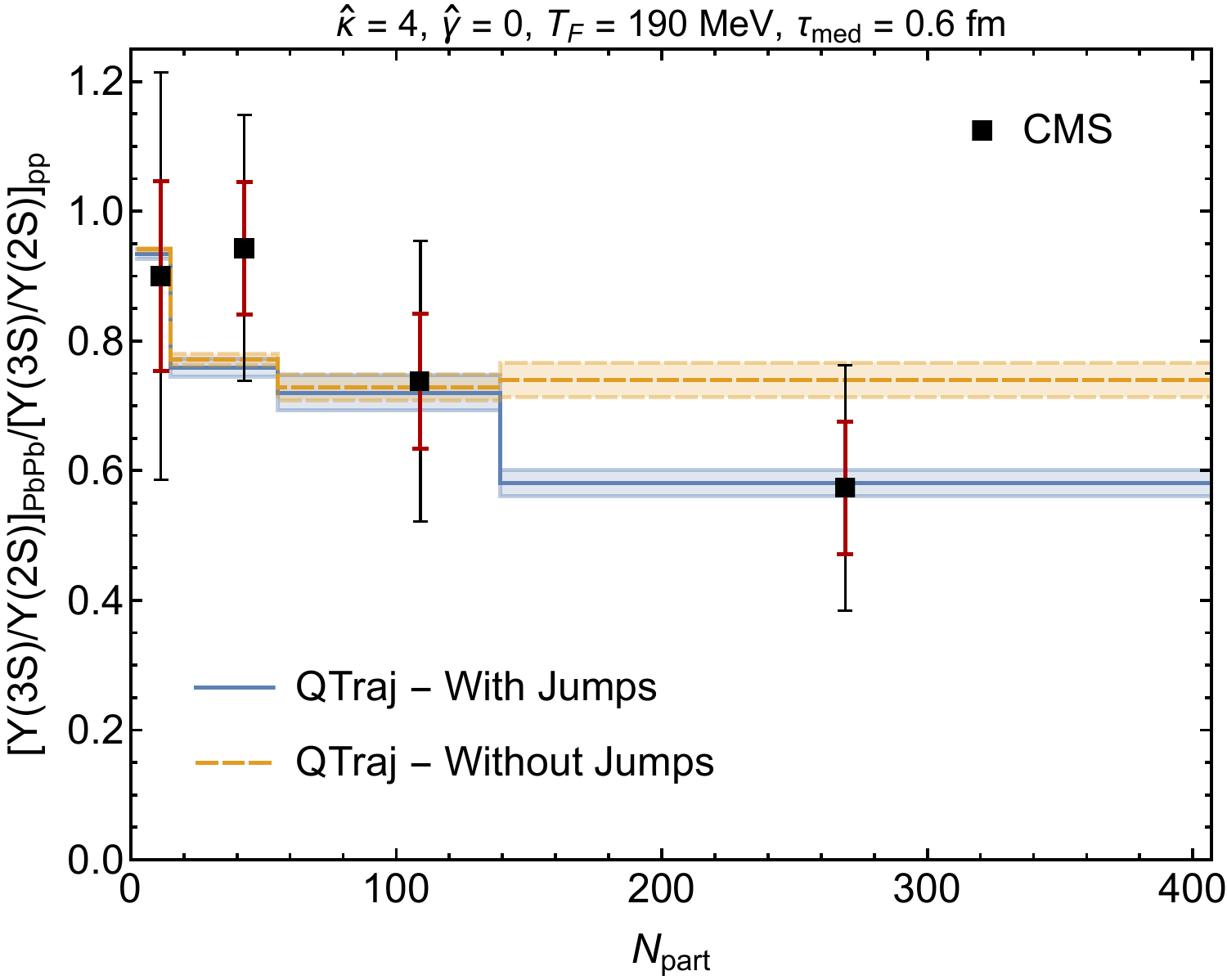}
    \vspace{-4mm}
    \caption{The 3S/2S double ratio as a function of $N_{\rm part}$. Line styles are the same as Fig.~\ref{fig:ratio2s1svsnpart}.  The centrality classes used were 0-30, 30-50, 50-70, and 70-90\%.   Experimental data are from Ref.~\cite{CMS-PAS-HIN-21-007}. }
    \label{fig:ratio3s2svsnpart}
\end{figure}

In Figs.~\ref{fig:ratio2s1svsnpart}{\,-\,}\ref{fig:ratio3s2svsnpart}, we compare our predictions for the 2S to 1S, 3S to 1S, and 3S to 2S double ratios defined via $(n_{AA}[nS]/n_{AA}[n'S])/(n_{pp}[nS]/n_{pp}[n'S])$ as a function of $N_{\rm part}$.
In the case of the CMS data, for the 2S to 1S double ratio, we inferred this observable by taking the ratio of the reported $R_{AA}[nS]$ to $R_{AA}[n'S]$ data as no update of the double ratio was reported in Ref.~\cite{CMS-PAS-HIN-21-007}.  
As a result, for this data set, we do not report the statistical and systematic uncertainties separately.  
For the ATLAS 2S to 1S double ratio, the black and red error bars correspond to statistical and systematic uncertainties, respectively.  
In the case of the CMS 3S to 1S double ratio, we inferred this observable from the 3S to 2S double ratio reported by CMS (shown in Fig.~\ref{fig:ratio3s2svsnpart}) and the inferred 2S to 1S double ratio presented in Fig.~\ref{fig:ratio2s1svsnpart}.  
For the ATLAS 3S to 1S double ratio, we use their reported combined 2S+3S to 1S double ratio~\cite{Sirunyan:2018nsz}. 
As Figs.~\ref{fig:ratio2s1svsnpart}{\,-\,}\ref{fig:ratio3s2svsnpart} demonstrate, this observable is only well explained when quantum regeneration is included.  
Without quantum regeneration, the 2S to 1S and 3S to 1S double ratios are under predicted and the 3S to 2S double ratio is slightly over predicted.

\begin{figure}[t]
    \centering
    \includegraphics[width=0.98\linewidth]{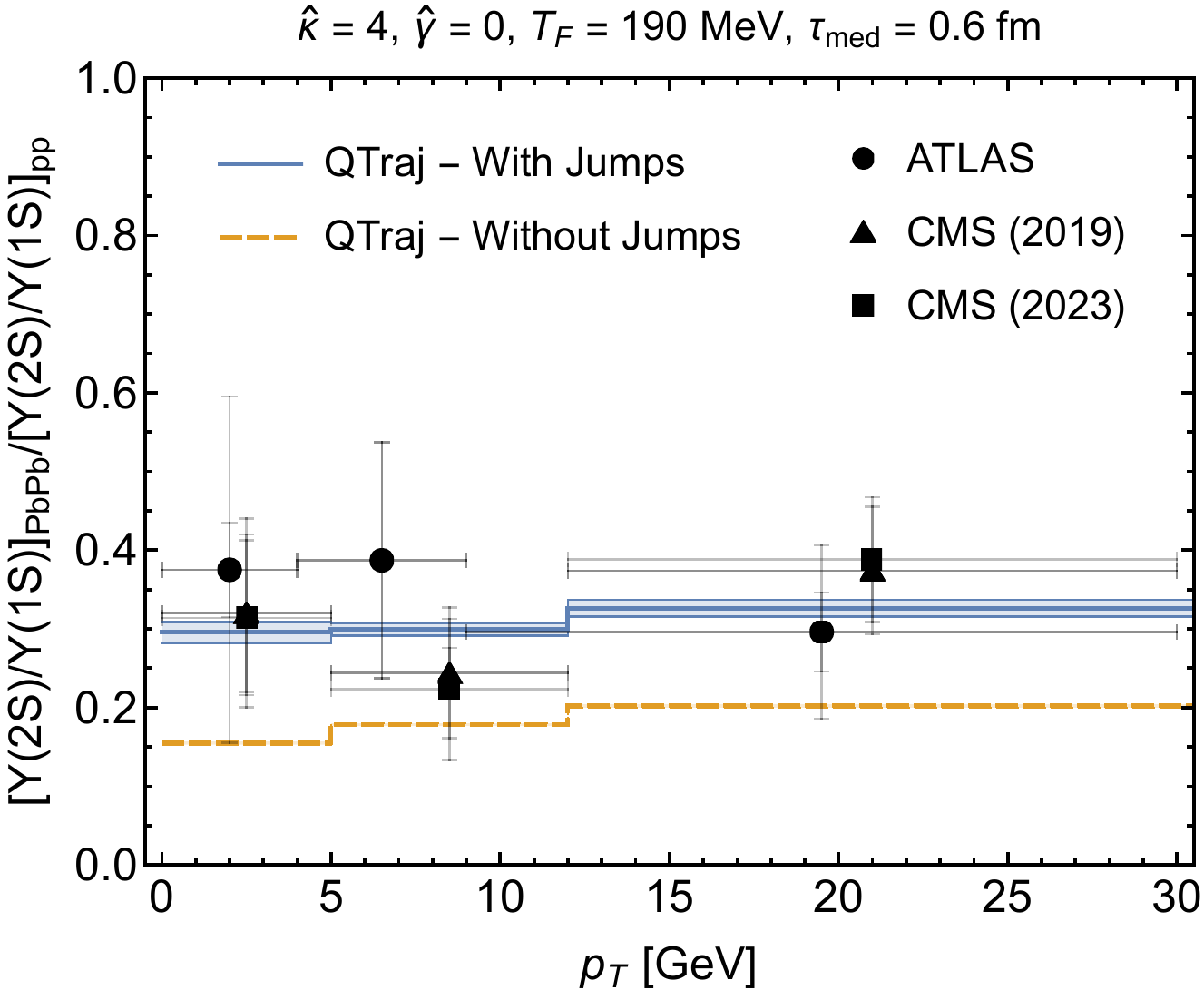}
    \vspace{-3mm}
    \caption{The 2S/1S double ratio as a function of $p_T$. Line styles are the same as Fig.~\ref{fig:ratio2s1svsnpart}. The experimental data were inferred from the results of Refs.~\cite{Sirunyan:2018nsz,CMS-PAS-HIN-21-007}.  }
    \label{fig:ratio2s1svspt}
\end{figure}

\begin{figure}[t]
    \centering
    \includegraphics[width=0.98\linewidth]{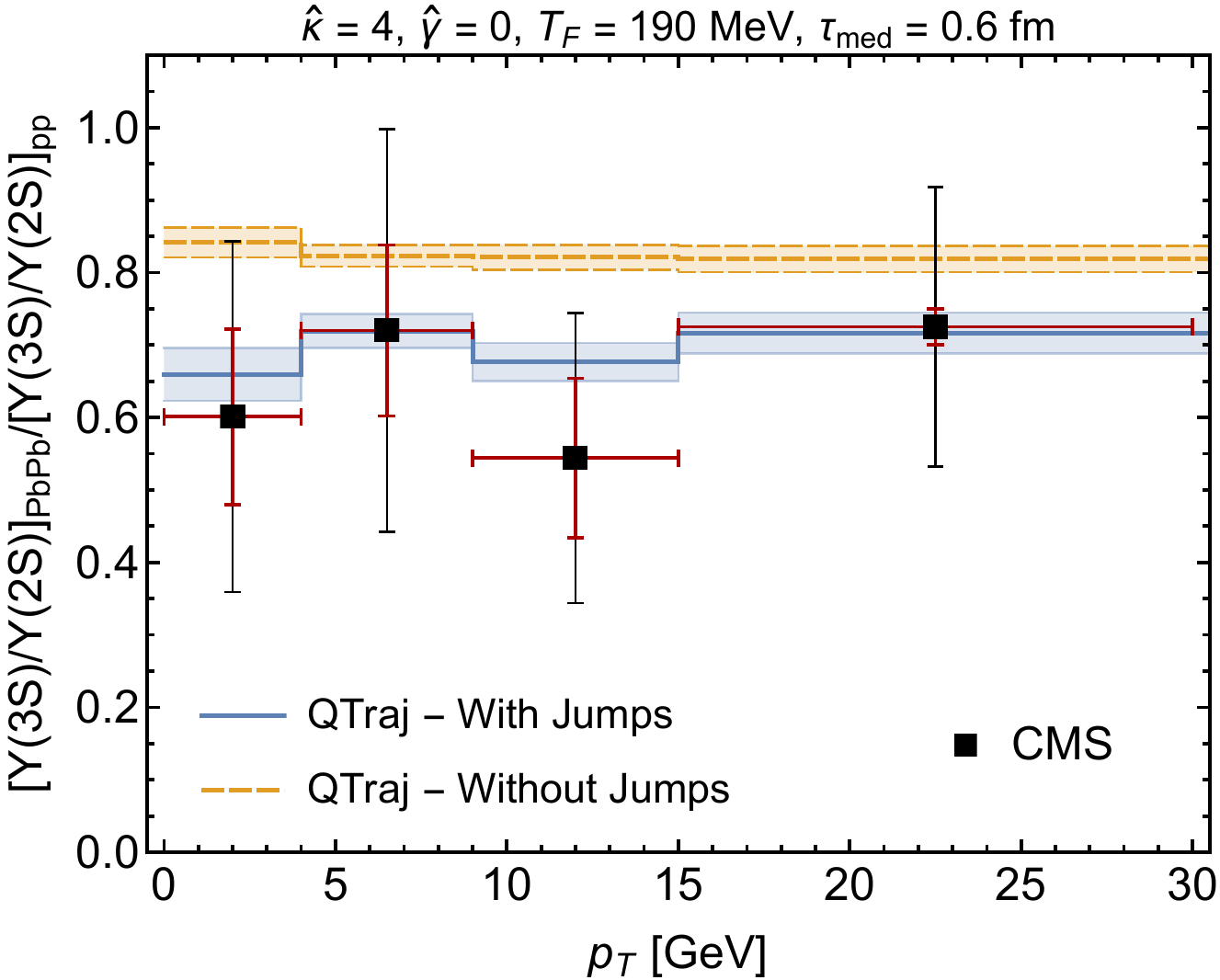}
    \vspace{-2mm}
    \caption{The 3S/2S double ratio as a function of $p_T$. Line styles are the same as Fig.~\ref{fig:ratio2s1svsnpart}. The experimental data are from Ref.~\cite{CMS-PAS-HIN-21-007}. }
    \label{fig:ratio3s2svspt}
\end{figure}

Finally, in Figs.~\ref{fig:ratio2s1svspt} and \ref{fig:ratio3s2svspt}, we present predictions for the 2S to 1S and 3S to 2S double ratios as a function of $p_T$.  
The data for the 2S to 1S double ratio labeled  `CMS (2023)' were inferred from the reported $p_T$ dependence of $R_{AA}[1S]$ and $R_{AA}[2S]$.  
For the ATLAS and `CMS (2019)' data, the Collaborations reported their computed 2S to 1S double ratio directly.  
As can be seen from these two figures, without jumps, these double ratios are in poorer agreement with the data, again indicating the importance of quantum regeneration for describing the suppression of excited states in heavy-ion collisions.  
We note, however, that both with and without jumps our NLO QTraj simulations predict a weak dependence of these double ratios on $p_T$.

{\em Conclusions} -- In this Letter we have shown that the inclusion of quantum regeneration allows for a quite satisfactory description of the world's collected data on suppression of bottomonium ground and excited states in a QGP. 
Leaving out quantum regeneration considerably drifts predictions away from the data.
These results confirm in a fully quantum and rigorous QCD setting semiclassical studies of singlet-octet transitions \cite{Blaizot:2018oev,Yao:2018nmy,Yao:2020xzw,Yao:2020eqy,Miura:2022arv}.
We can thus evolve the singlet and octet density matrix considering  the effect of medium-induced color and angular momentum transitions between the states.  
Looking to the future, it will be interesting to examine the effect of temperature dependence in the heavy-quarkonium transport coefficients, the effect of fluctuating hydrodynamical backgrounds \cite{Alalawi:2022gul}, and the possibility of further improvements to the underlying pNRQCD treatment.

\acknowledgements{
{\em Acknowledgements} --
N.B. and A.V. acknowledge support by the DFG cluster of excellence ORIGINS funded by the Deutsche Forschungsgemeinschaft under Germany's Excellence Strategy - EXC-2094-390783311.
M.\'{A}.E was supported by European Research Council project ERC-2018-ADG-835105 YoctoLHC, by the Maria de Maetzu excellence program under projects CEX2020-001035-M and CEX2019-000918-M, the Spanish Research State Agency under projects PID2020-119632GB-I00 and PID2019-105614GB-C21, the Xunta de Galicia (Centro singular de investigaci\'on de Galicia accreditation 2019-2022; European Union ERDF).
M.S., A.I., and A.T. were supported by U.S. Department of Energy Award No.~DE-SC0013470 and National Science Foundation Award No.~2004571.
M.S. also thanks the Ohio Supercomputer Center under the auspices of Project No.~PGS0253.  
P.V.G. was supported by the U.S. Department of Energy
%, Office of Science, Office of High Energy Physics, under 
Award No.~DE-SC0019095.
P.V.G. is grateful for the support and hospitality of the Fermilab theory group.
Fermilab is operated by Fermi Research Alliance, LLC under Contract No.~DE-AC02-07CH11359 with the United States Department of Energy.
}

\appendix

\begin{widetext}

\section*{Supplemental material}

In this supplement, we provide details concerning our parameter scan.  We considered sixteen combinations, varying $\hat\kappa \in \{2,3,4,5\}$ and $\hat\gamma \in \{-3.5, -2.6, 0, 1\}$.  For each of the sixteen combinations, we computed the quadratic differences between our next-to-leading order result for $R_{AA}[1S,2S]$ as a function of $N_{\rm part}$ and the central values of the corresponding experimental data.  We did not include $R_{AA}[3S]$ in the analysis since only one experiment has reported data for $R_{AA}[3S]$.  For each experiment $i \in \{\text{ALICE}, \text{ATLAS}, \text{CMS} \}$ and state $j \in \{ \text{1S}, \text{2S} \}$, we computed the distance measure
$$
\mes^2(i,j) \equiv \frac{1}{N_{\rm data}^i} \sum_{k=1}^{N_{\rm data}^i} \left( \frac{R_{AA}^{\rm QTraj}[j,k] - R_{AA}^i[j,k]}{R_{AA}^i[j,k]} \right)^2 ,
$$
where $k$ indexes the experimental data points in $N_{\rm part}$.  At each point, we took the central value of the experimental data points and QTraj predictions, with the QTraj predictions being linearly interpolated in $N_{\rm part}$ to match the precise values of $N_{\rm part}$ reported by each experiment.  We then constructed an average over experiments for each state $\mes^2(j) \equiv \frac{1}{3} \sum_i \mes^2(i,j)$ and a combined measure $\mes^2 \equiv \sum_j \mes^2(j)$.
These measures are shown in Fig.~\ref{fig:fitPlot}.  The top and bottom rows show the results obtained with and without jumps (quantum regeneration), respectively.  To generate these figures and the conclusions reported next, the values of $\mes^2(j)$ and $\mes^2$ were interpolated using cubic and quadratic interpolations in $\hat\kappa$ and $\hat\gamma$, respectively.  
\begin{figure}[b]
    \centering
    \includegraphics[width=0.975\linewidth]{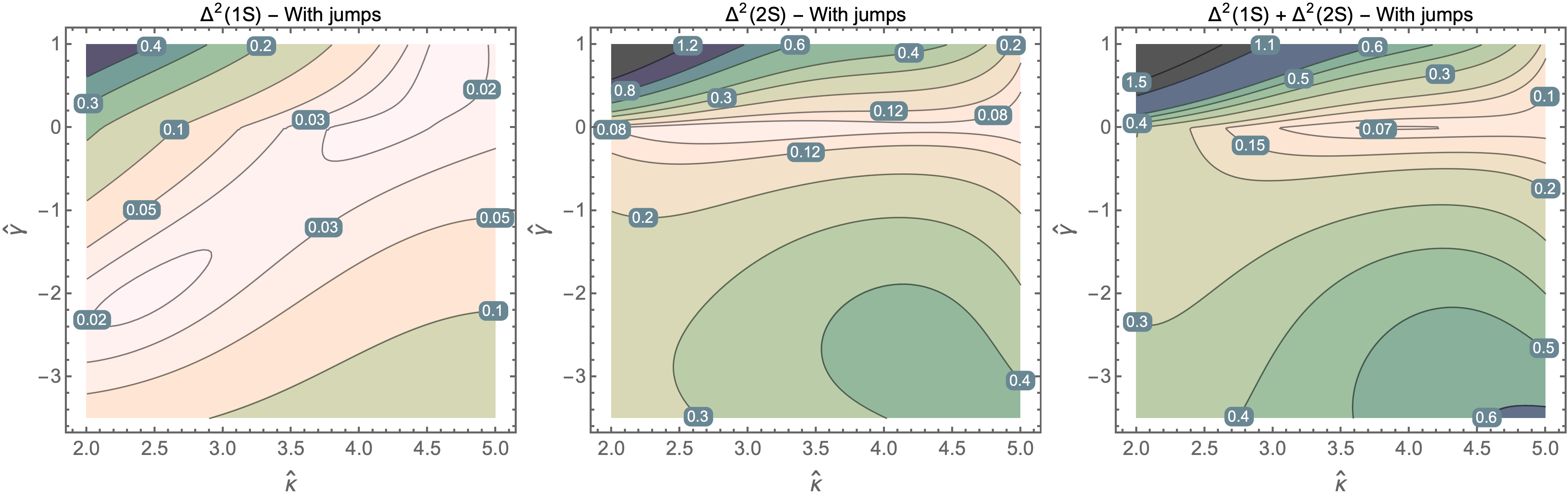}\\[0.5em]
     \includegraphics[width=0.975\linewidth]{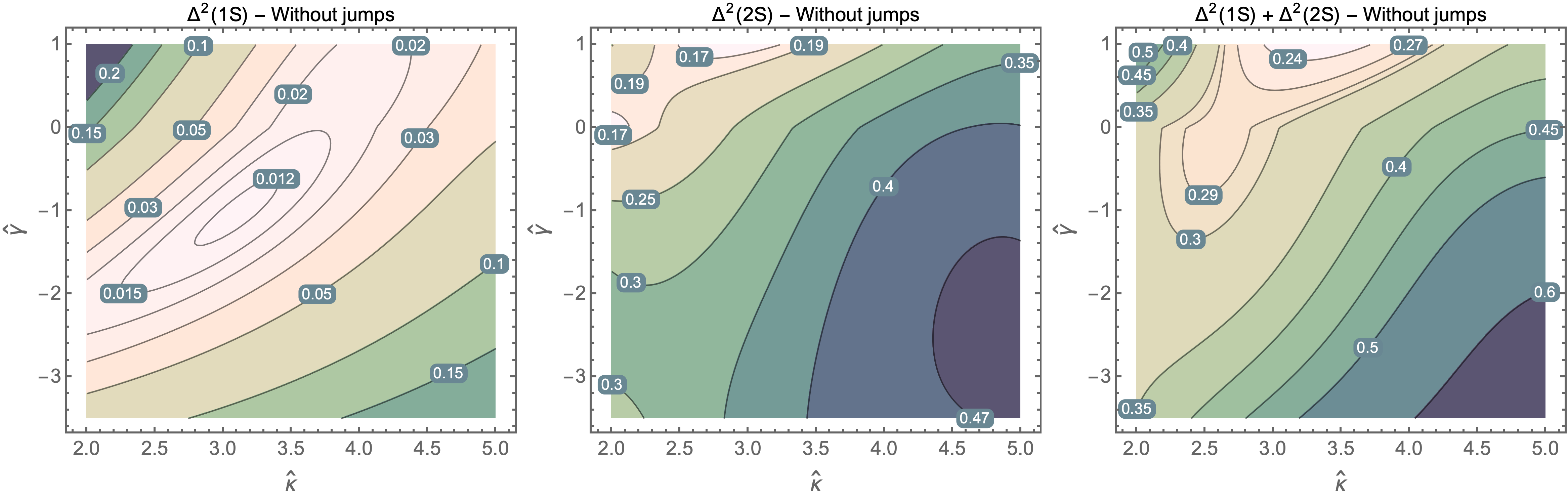}
    \vspace{-2mm}
    \caption{Theory and experiment distance measures obtained with jumps (top row) and without jumps (bottom row).}
    \label{fig:fitPlot}
\end{figure}
When including jumps, we found that there was a reasonable overlap in the best descriptions for the 1S and 2S states and a minimum in the combined $\mes^2$ of $\mes^2_\text{min} = 0.065$ at $\hat\kappa=3.9$ and $\hat\gamma=0$.  Without jumps, we found that the 1S and 2S best description regions do not overlap, with the 2S description being best when $\hat\kappa \sim 2.8$ and $\hat\gamma=1$, which results in a poor description of $R_{AA}[1S]$ as a function of $N_{\rm part}$.  Without jumps, the minimal value of the combined $\mes^2$ was $\mes^2_\text{min} = 0.22$ at $\hat\kappa=3.3$ and $\hat\gamma=1$.  The resulting best description without jumps does not describe well the suppression of either the 1S or 2S states.  Based on these findings, in the main body of the text we compare results obtained from the case $\hat\kappa=4$ and $\hat\gamma=0$, which are the values closest to the interpolated minimum for $\mes^2$ when including quantum jumps.

\end{widetext}

%%%%%%%%%%%%%%%%%%%%%%%%%%%%%%%%%%%%%%%%%%%%%%%%
\bibliography{qtraj-nlo-jumps}
%%%%%%%%%%%%%%%%%%%%%%%%%%%%%%%%%%%%%%%%%%%%%%%%

\end{document}